# Efficient Terahertz Generation Using Fe/Pt Spintronic Emitters Pumped at Different Wavelengths

Evangelos Th. Papaioannou[1], Garik Torosyan[2], Sascha Keller[1], Laura Scheuer[1], Marco Battiato[3], Valynn Katrine Mag-usara[4], Johannes L'huillier[2], Masahiko Tani[4], René Beigang[1]

[1]Department of Physics and Research Center OPTIMAS, University of Kaiserslautern, Kaiserslautern 67663, Germany
[2]Photonic Center Kaiserslautern, Kaiserslautern 67663, Germany
[3]School of Physical and Mathematical Sciences, Nanyang Technological University, Singapore 637371, Singapore
[4]Research Center for Development of Far-Infrared Region, University of Fukui, Japan

**Recent studies in spintronics have highlighted ultrathin magnetic metallic multilayers as a novel and very promising class of broadband terahertz radiation sources. Such spintronic multilayers consist of ferromagnetic (FM) and non-magnetic (NM) thin films. When triggered by ultrafast laser pulses, they generate pulsed THz radiation due to the inverse spin-Hall effect – a mechanism that converts optically driven spin currents from the magnetized FM layer into transient transverse charge currents in the NM layer, resulting in THz emission. As THz emitters, FM/NM multilayers have been intensively investigated so far only at 800-nm excitation wavelength using femtosecond Ti:sapphire lasers. In this work, we demonstrate that an optimized spintronic bilayer structure of 2-nm Fe and 3-nm Pt grown on 500 μm MgO substrate is just as effective as a THz radiation source when excited either at λ = 800 nm or at λ = 1550 nm by ultrafast laser pulses from a fs fiber laser (pulse width ~100 fs, repetition rate ~100 MHz). Even with low incident power levels, the Fe/Pt spintronic emitter exhibits efficient generation of THz radiation at both excitation wavelengths. The efficient THz emitter operation at 1550 nm facilitates the integration of such spintronic emitters in THz systems driven by relatively low cost and compact fs fiber lasers without the need for frequency conversion.**

*Index Terms*— Epitaxial bilayers, fs lasers, Inverse spin Hall effect, THz spintronics.

## I. INTRODUCTION

THE USE OF BILAYERS composed of a ferromagnetic metal (FM) layer attached to a non-magnetic metal (NM) as sources for THz emission opens a new direction in physics. It combines the field of THz with the field of spintronics, since the creation of the THz pulses is based on the generation of spin currents [1]. The spin current is created in the magnetic layer and then it diffuses through a super-diffusive process [2, 3] in the non-magnetic metal. Subsequently, via the inverse spin Hall effect (ISHE) it is converted into a transient charge current that gives rise to THz radiation. Previous investigations have nicely shown this effect in different FM/NM systems [3]-[8]. Furthermore, in our recent work [8] we have optimized the spintronic emitters based on epitaxial grown Fe/Pt layers: the emitter was optimized with respect to layer thickness, growth parameters, substrates and geometrical arrangement. In addition, the experimentally determined optimum layer thicknesses for Fe and Pt were in qualitative agreement with simulations of the spin current induced in the ferromagnetic layer. The model in [8] describes the dependence of the THz emission on the individual layer thicknesses and not only on the total layer thickness [2] of Fe and Pt layers. The model explains the onset of THz generation above a certain Fe layer thickness and the maximum at relatively small layer thicknesses. It takes into account all successive effects after the laser pulse strikes on the bilayer, including the absorption of the fs pulse in the Pt and Fe layers, the generation and diffusion of spin currents in Fe and Pt as well as the generation of THz radiation and its attenuation in the metal layers.

The technological interest in THz spintronic emitters is large since spintronic emitters are efficient, easy-to-use, robust and need no electrical connections or lithographic structuring [3 - 8]. The capabilities of such metallic thin-film stacks in applications as a source of intense and broadband terahertz electromagnetic fields where recently investigated [9]. There, a W/CoFeB/Pt trilayer grown on a large-area glass substrate (diameter of 7.5 cm) was excited. In spite of the great technological perspectives and the many interesting results published up to date, the FM/NM multilayers have been intensively investigated so far only at 800 nm excitation wavelength using femtosecond Ti:sapphire lasers [1-9]. Excitation schemes with other wavelengths could be of importance concerning the future use of such THz sources. Here, the THz emission is probed with a pump laser at 1550 nm and compared to the corresponding properties at 800 nm excitation.

Corresponding author: Evangelos Th. Papaioannou (e-mail: epapa@physik.uni-kl.de).

## II. EXPERIMENT

### A. Experimental details

A standard terahertz time-domain spectroscopy (THz-TDS) system, described in detail in [8, 10] has been used for generation and measurements of THz waveforms from different spintronic emitters (see Fig 1). The system is driven by a femtosecond Ti:Sa laser delivering sub-100 fs optical pulses at a repetition rate of 75 MHz with an average output power of typically 600 mW. Alternatively, a fiber laser at 1.55 µm wavelength with a repetition rate of 100 MHz and an output power of up to 50 mW was used. The laser beam is split into a pump and probe beam by a 90:10 beam-splitter. The stronger part is led through a mechanical computer-controlled delay line to pump the THz emitter, and the weaker part is used to gate the detector photoconductive switch with a 20 µm dipole antenna. The PCA emitter used in a standard TDS set-up is replaced by a spintronic (ST) Fe/Pt bilayer sample. The sample is mounted perpendicular to the direction of the pump beam. A weak external magnetic (20 mT) field is applied perpendicular to the pump beam to saturate the magnetization in the direction of the easy axis of Fe. The direction of the magnetic field determines the polarization of the THz field which is always perpendicular to the direction of the pump beam and the direction of the magnetic field. The PCA detector is sensitive to the polarization direction of the incident THz wave and, therefore, the polarization direction has to be stabilized by the external magnetic field and the orientation of the easy axis of the Fe/Pt emitter.

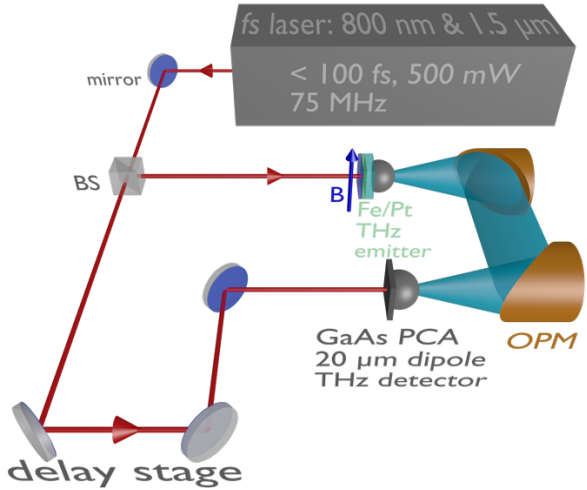

**Figure 1:** THz time-domain emission spectroscopy setup (THz-TDES) was used to measure the THz emission of the optimized Fe/Pt spintronic emitter.

The optical pump beam is sharply focused with an aspherical short-focus lens down to a spot diameter of less than 10 µm in the plane of the metal layers. It excites spin polarized electrons in the magnetic layer (Fe) which give rise to a spin-current, which in turn excites a transverse transient electric current in the Pt layer. The current transient results in THz pulse generation of sub-picosecond duration with a center wavelength around 300 µm being emitted forward and backward into free space in the form of a strongly divergent beam. A hyperhemispherical Si-lens being attached to the back side of the sample collects it into a divergent beam in the form of a cone and directs it further to an off-axis parabolic mirror for collimation. With the lens attached an enhancement factor of up to 30 in electric field amplitude has been observed. The function of the Si-lens is to reduce the angular dimensions of the emitted THz wave already at its source. Since the central wavelength of the THz emission is about 300 µm and is much bigger than dimensions of its source (about 10µm in diameter), the beam, if not being collected with the help of an additional optical element (in our case- Si-lens), tends to diffract immediately and fill the half space behind the emitter sample. Without the Si-lens most of THz beam would pass by the first parabolic mirror and correspondingly not reach the detector, reducing essentially the efficiency of the THz optics. But since the comparing measurements are made with the same Si-lens, only the differences (if any for different pump wavelengths) are of interest. Another and identical off-axis parabolic mirror in the reversed configuration directs the collimated THz beam onto the second Si lens which finally focuses the beam through the GaAs substrate of the detector PCA onto the dipole gap for detection. In this way the THz optical system consisting of the two Si-lenses and the two parabolic mirrors images the point source of THz wave on the emitter surface onto the gap of the detector PCA and ensures an efficient transfer of the emitted THz emission from its source to the detector.

In our set-up the THz beam path is determined by the silicon lens on the emitter, the parabolic mirrors and the silicon lens on the photoconductive antenna of the detector. The alignment of these components is not changed during an exchange of the spintronic emitter. If in addition the position of the pump beam focus remains constant the spintronic emitter can easily be exchanged without changing the beam path as the lateral position of the focus on the emitter is not critical assuming a homogeneous lateral layer structure.

The frequency response of the photoconductive dipole antenna with 20 µm dipole length limits the observable bandwidth. With this dipole length a maximum in the detector response around 1 THz can be expected with a reduction to 50% at 330 GHz and 2 THz. The 10% values are at 100 GHz and 3 THz. Above 3 THz the frequency response of the detector is very flat and at 8 THz a strong phonon resonance in GaAs which is used as substrate material for the photoconductive antenna causes strong absorption of the THz radiation. Above 8 THz no THz radiation was detected.

The delay line provides for the synchronous arrival of the weaker part of the optical pulse and that of the THz pulse at the detector antenna gap from either side. In each position of the delay line the transient current induced in the detector by the electric field of the THz sub-picosecond pulse is proportional to its instantaneous electrical field value. By scanning the "open" state of the detector in time the THz pulse shape can be sampled. For our measurements we have used a scan range of 33 ps with a step width of 0.4 µm of our delay line, a scan speed of 30 µm/s and an integration time of 100 ms.

An important characteristic of the ST emitter is the mean power of the generated THz radiation. We have measured it directly using a thin-film pyroelectric THz detector of model THz20 in combination with the current preamplifier CPA 119 calibrated together at the PTB, Braunschweig-Berlin [11]. The pump beam for using the detector was mechanically chopped at 20 Hz, and the output signal in the form of quasi-rectangular pulses measured on an oscilloscope. The peak-to-peak amplitude of the measured pulses is proportional to the average THz power, with the calibrated sensitivity of 154.4 V/W. Considering the losses of the THz power through the Si lens, we obtain an estimated absolute average power of almost 50 µW for pump power of 500 mW with 800 nm wavelength.

We deposited a series of Fe/Pt bilayer and Fe/Cu/Pt trilayers varying the thickness t of Fe, Pt, and Cu. The samples were grown by electron beam evaporation in an ultrahigh vacuum chamber. Fe films were grown epitaxially on 0.5 mm-thick MgO (100) substrates. The Pt layer follows the Fe and grows epitaxially by rotating its fcc cell by 45 degrees with respect to the Fe lattice [12-14]. Typical dimensions of the samples were 1 cm × 1 cm.

### III. RESULTS AND DISCUSSION

#### A. Probing the THz emission through a Cu interlayer

To demonstrate that the ISHE is responsible for the THz generation we have added a Cu layer between Fe and Pt. In this way, we separate the spin injector (Fe) and the spin sink (Pt) layer. The Cu interlayer also excludes any direct exchange coupling between Fe and Pt, the so-called magnetic proximity effect where a non-magnetic metal obtains a finite magnetic moment [15]. The magnetic proximity effect can influence the injection of spin current as it has recently been discussed in the spin pumping experiments [16].

In Figure 2 we compare the THz pulses of two samples: that of an optimized Fe (2 nm) / Pt (3 nm) bilayer with that of a Cu interlayer Fe (2 nm) /Cu (3 nm) /Pt (3 nm) system. Cu is a non-magnetic conductor with long spin diffusion length. The effective spin current emitted from the ferromagnetic Fe layers travels through the Cu layer and the spin current that reaches the Pt layer is usually smaller than the initial one. That is the reason for the lower amplitude of the THz pulse for the interlayer case. However, the insertion of a thin Cu layer between Fe and Pt retains the spin current because the Cu thickness is much smaller than its own spin diffusion length something that allows the trilayer to give rise to strong THz emission. The down factor of 2 can be attributed to the reduction of the interface transmission due to the Cu layer. Similar downscaling of the signal was observed in spin pumping ferromagnetic resonance (FMR) experiments in Co/Pt and Co/Cu/Pt systems [17].

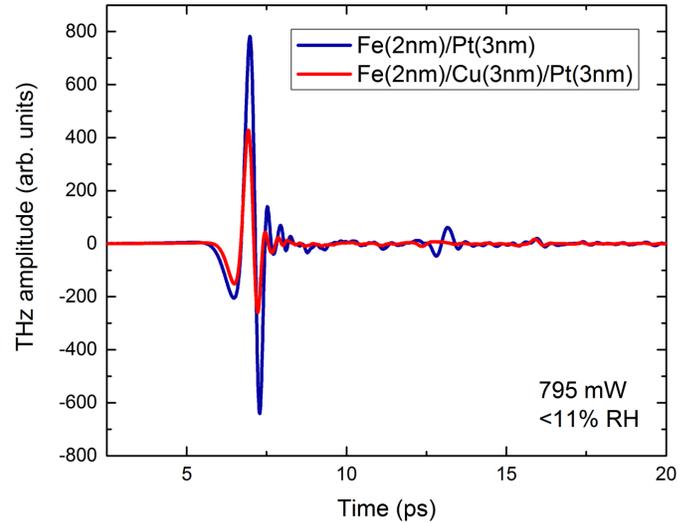

*Figure 2: Probing the spin current transport as the source of the THz generation: comparison of the THz emission with the presence or not of a Cu interlayer of 3 nm thickness.*

#### B. Probing the THz emission with the wavelength of the pump laser

Figure 3 shows the waveforms and spectra of the THz radiation when the optimized Fe/Pt bilayer structure was used as emitter in a standard THz time-domain spectroscopy setup. With a time domain set-up the voltage measured at the photoconductive antenna as a function of delay time between emitter and detector pulse is always directly proportional to the electric field of the THz pulse [10]. A Fourier transform leads to the spectral amplitude as a function of frequency as amplitude and phase are known in the time domain. The data in Fig. 3 refer to an optimized spintronic bilayer structure consisting of 2-nm Fe and 3-nm Pt grown on 0.5 mm-thick MgO substrate and measured with two different wavelength pump excitation laser sources, λ = 800, 1550 nm. The emitter is as effective as a THz radiation source when excited either at 800 nm or at 1550 nm. The ultrafast laser pulses of 1550 nm are driven from a relatively low cost and compact fs fiber laser (pulse width ~100 fs, repetition rate ~100 MHz). Even with low incident power levels of 1 mW and 7 mW, the Fe/Pt spintronic emitter exhibits efficient generation of THz radiation at different excitation wavelengths.

Figure 3 proves that the excitation of out-of-equilibrium spin polarized electrons (responsible for the generation of spin current and the subsequent THz emission) with less energy (λ = 1550 nm, E = 0.826 eV) is as effective as the use of higher excitation energy (λ = 800 nm, E = 1.55 eV).

One would expect that the higher the energy of the electrons and holes the bigger is the asymmetry of the two spin channels since they are located in different bands, with clearly different group velocities.

However, we observe through the emitted THz pulses that the spin current generation at lower energy is as efficient as for higher energy. In order to understand this behavior, we have to take into account that the spin diffusion requires not only the asymmetry between the two spin channels, but also between electrons and holes.

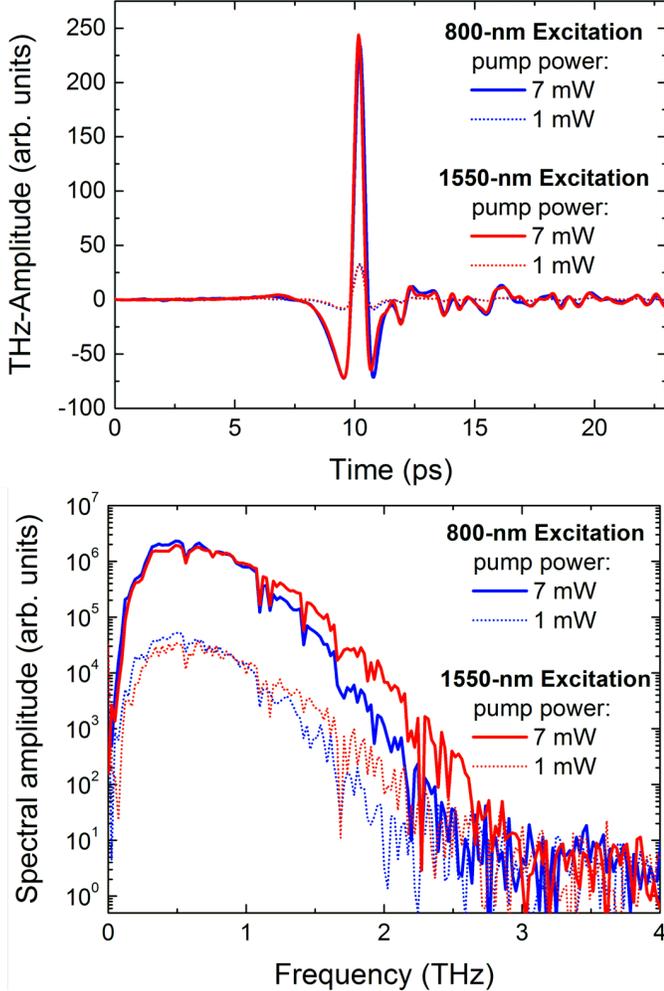

*Figure 3: Waveforms and spectra of the THz radiation obtained with two pump laser wavelengths (800 and 1550 nm) for an optimized Fe/Pt bilayer structure that was used as emitter in a standard THz time-domain emission spectroscopy setup.*

The reason behind the observed effect is a consequence of the fact that the out of equilibrium transport is done not only by the electrons directly excited by the laser, but also by all the electrons that are excited at intermediate energies due to the scattering of the first-generation electrons. We remind the reader that electron-electron scattering lifetimes sharply decrease with the energy of the excited electron. This means that the direct impact of high energetic electrons on the transport is minor. Instead, the most important impact is that by scattering with another electron the electrons will lose energy, they will go down to an intermediate energy, and transfer that energy to secondary electrons. This results into an intermediate energy electrons multiplication, which is very similar to a direct excitation by a laser at lower frequency. Even with lower energies now, they will have longer lifetimes and they will contribute to the transport more importantly.

The argument above, however, holds only for medium to high frequency excitations. If we were to excite the system with very low frequency photons, other effects will become important, changing the qualitative scenario proposed above. We will have to take into account that the spin diffusion requires not only Therefore, in case of very low frequency excitation, even if the number of excited electrons at very low energies (we assume 0.1 eV above the Fermi energy) is very high, and they are expected to contribute importantly to the transport, the similarity of their transport with those of their respective holes leads to an almost complete cancellation of spin transport (we remind that electrons and holes within the same spin channel carry opposite spin). If an electron and a hole in the same spin channel travel in the same way, it will create no charge or spin current it will only displace energy.

Figure 3 shows that the important factor is the total energy that it is originally deposited into the system. The THz spectra for total injected power of 1 mW and 7 mW look very similar, however the intensity is higher for the higher power. Figure 4 finally shows the linear dependence of the THz electric field amplitude with respect to the total laser power for both wavelengths. Figure 4 has been obtained as a function of pump power for a fixed spot size of the pump beam. With a spot size of 5 μm radius, a repetition rate of 76 MHz of our pump laser we changed the pump power on the sample from 1 mW to 20 mW without any damage of the sample.

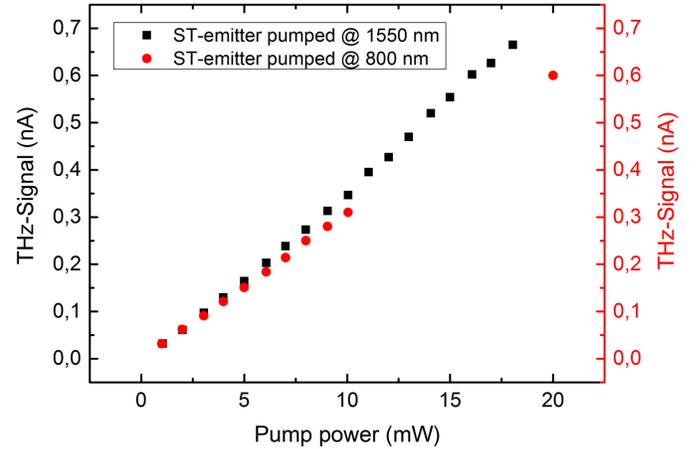

*Figure 4: Maximum E-field amplitude as a function of pump power of a laser with two pump wavelengths: black squares refer to 1550 nm wavelength and red circles to 800 nm measured with different detector PCAs.*

### IV. CONCLUSION

We have shown that Fe/Pt spintronic THz emitter is as efficient operated at λ = 1550 nm excitation wavelength as when excited with λ = 800 nm. This fact opens up the feasibility of directly integrating spintronic heterostructures in THz systems driven by relatively low cost and compact fs fiber lasers without the need for frequency conversion.

### ACKNOWLEDGMENT

E.Th.P., S.K., L.S thank the Deutsche Forschungsgemeinschaft (DFG) for the support through the collaborative research center SFB TRR 173: SPIN+X Project B07 and the Carl Zeiss Foundation.